\documentclass[a4paper,12pt, epsfig]{article}\def\pr{0}\def\letter{0}

\usepackage{epsfig}
\usepackage{epstopdf}
\usepackage{graphicx}
\usepackage{ifthen}

\usepackage{amsmath}
\usepackage[psamsfonts]{amssymb}
\usepackage{euscript}

\usepackage{latexsym}
\usepackage[arrow,matrix,curve]{xy}

\newskip\humongous \humongous=0pt plus 1000pt minus 1000pt

\newif\ifdtup

\def\,{\hspace{-.1cm}}
\def\hsp{,\hspace{.7cm}}

\def\fc#1#2 {\frac{n}{q}#1\frac{n}{q}#2}

\def\kt{\kappa}
\def\kt{\mathfrak{K}}
\def\kt{\mathrm{S}}
\def\ks{|\kt\rangle}
\def\bdk{B^\dag_\kt}

\newcommand{\vac}{\ensuremath{|0\rangle}}

\renewcommand{\sin}{\textrm{sin}}

\renewcommand{\tanh}{\textrm{tanh}}
\newcommand{\sech}{\textrm{sech}}
\newcommand{\csch}{\textrm{csch}}

\def\cc{\mathcal{C}}

\renewcommand{\theequation}{\arabic{section}.\arabic{equation}}
\renewcommand{\(}{\begin{equation}}
\renewcommand{\)}{end{equation} \vspace{-.05in}\linebreak}

\newcounter{saveeqn}
\newcounter{savealpheqn}

\newcommand{\alpheqn}{\setcounter{saveeqn}{\value{equation}}%
  \stepcounter{saveeqn}\setcounter{equation}{0}%
  \renewcommand{\theequation}{\mbox{\arabic{section}.\arabic{saveeqn}
\alph{equation}}}
  \renewcommand{\)}{\end{equation}}}
\def\part#1{\frac{\partial}{\partial{#1}}}%
\def\group#1{\refstepcounter{equation}\setcounter{saveeqn}
 {\value{equation}}%
  \label{#1}\setcounter{equation}{0}%
\renewcommand{\theequation}{\mbox{\arabic{section}.\arabic{saveeqn}
\alph{equation}}}
  \renewcommand{\)}{\end{equation}}}
\newcommand{\reseteqn}{\setcounter{equation}{\value{saveeqn}}%
  \renewcommand{\theequation}{\arabic{section}.\arabic{equation}}%
  \renewcommand{\)}{\end{equation}}}

\newcommand{\aalpheqn}{\setcounter{saveeqn}{\value{equation}}%
  \stepcounter{saveeqn}\setcounter{equation}{0}%
  \renewcommand{\theequation}{\mbox{
        \Alph{subsection}.\arabic{saveeqn}\alph{equation}}}
   \renewcommand{\)}{\end{equation}}}
\newcommand{\areseteqn}{\setcounter{equation}{\value{saveeqn}}%
  \renewcommand{\theequation}{\Alph{subsection}.\arabic{equation}}%
  \renewcommand{\)}{\end{equation}}}

\renewcommand{\thefootnote}{\alph{footnote}}
\renewcommand{\(}{\begin{equation}}
\renewcommand{\)}{\end{equation}}
\newcommand{\ba}{\begin{eqnarray}}
\newcommand{\ea}{\end{eqnarray}}
\renewcommand{\a}{\alpha}
\renewcommand{\b}{\beta}

\newcommand{\cbp}{\mathop{\vtop{\ialign{##\crcr
   $\hfil\displaystyle{}\hfil$\crcr\noalign{\kern-13pt\nointerlineskip}
   \BIG{)}\hskip0pt\crcr\noalign{\kern3pt}}}}}
\newcommand{\pa}{\mathop{\vtop{\ialign{##\crcr

$\hfil\displaystyle{\oplus}\hfil$\crcr\noalign{\kern+1pt\nointerlineskip
}
   \hspace{.08in}$^{\alpha=0}$\hskip6pt\crcr\noalign{\kern3pt}}}}}
\renewcommand{\hsp}{,\hspace{.3in}}
\newcommand{\p}{^\prime}
\newcommand{\pp}{^{\prime\prime}}

\newcommand{\Z}{\ensuremath{\mathbb Z}}

\def\rd{\rho^{\rm{}}}
\def\rv{\rho^{\rm{(vac)}}}
\def\rdiv{\rho^{\rm{(div)}}}
\def\rfin{\rho^{\rm{(fin)}}}
\def\rdif{\rho^{\rm{(dif)}}}
\def\qfin{Q_2^{\rm{(33fin)}}}

\catcode`\@=11
\def\vereq#1#2{\lower3pt\vbox{\baselineskip1.5pt \lineskip1.5pt
\ialign{$\m@th#1\hfill##\hfil$\crcr#2\crcr\sim\crcr}}}
\catcode`\@=12

\renewcommand{\(}{\begin{equation}}
\renewcommand{\)}{\end{equation}}

\def\mui#1{\mu^{(#1)}(\kt)}
\def\pin#1{\int \frac{d#1}{2\pi}}
\def\pink#1{\int \frac{d^{#1}k}{(2\pi)^{#1}}}

\def\pinkp#1{\int \frac{d^{#1}k\p}{(2\pi)^{#1}}}

\def\dpin#1{\int^+ \frac{d#1}{2\pi}}
\def\dpink#1{\int^+ \frac{d^{#1}k}{(2\pi)^{#1}}}

\def\Bd#1{B^\dag_{k_{#1}}}

\def\df{\mathcal{D}_f}
\def\B#1{B^\dag_{k_{#1}}}

\def\I{\mathcal{I}}
\def\vl#1#2#3{\vector(#1,#2){#3}\line(#1,#2){#3}}
\def\wk#1{\omega_{k_{#1}}}
\def\wkp#1{\omega_{k^{'}_{#1}}}

\newcommand{\beas}{\begin{eqnarray*}}
\newcommand{\eeas}{\end{eqnarray*}}

\newcommand{\bquo}{\begin{quote}}
\newcommand{\enqu}{\end{quote}}

\renewcommand{\Z}{{\mathbb Z}}

\def\ch{{\mathcal{H}}}

\def\ok#1{\omega_{k_{#1}}}
\def\okp#1{\omega_{k\p_{#1}}}
\def\V#1{V^{(#1)}[\sqrt{\lambda}f(x)]}
\def\ck{\csch\left(\frac{\pi k}{m}\right)}
\def\cks{\csch^2\left(\frac{\pi k}{m}\right)}
\def\sk{\sech\left(\frac{\pi k}{m}\right)}
\def\sks{\sech^2\left(\frac{\pi k}{m}\right)}
\def\kkk{\sum_{i}^3k_i}
\def\csk{\csch\left(\frac{\pi\kkk}{2\b}\right)}
\def\csks{\csch^2\left(\frac{\pi\kkk}{2\b}\right)}

\def\hp{{\hat{\Phi}}}

\newcommand{\beq}{\begin{equation}}
\newcommand{\eeq}{\end{equation}}
\newcommand{\bea}{\begin{eqnarray}}
\newcommand{\eea}{\end{eqnarray}}

\newskip\humongous \humongous=0pt plus 1000pt minus 1000pt

\newif\ifdtup

\catcode`\@=11

\ifthenelse{\equal{\letter}{0}}{ 

\@addtoreset{equation}{section}
\def\theequation{\arabic{section}.\arabic{equation}}

\def\@normalsize{\@setsize\normalsize{15pt}\xiipt\@xiipt
\abovedisplayskip 14pt plus3pt minus3pt%
\belowdisplayskip \abovedisplayskip
\abovedisplayshortskip \z@ plus3pt%
\belowdisplayshortskip 7pt plus3.5pt minus0pt}

\def\small{\@setsize\small{13.6pt}\xipt\@xipt
\abovedisplayskip 13pt plus3pt minus3pt%
\belowdisplayskip \abovedisplayskip
\abovedisplayshortskip \z@ plus3pt%
\belowdisplayshortskip 7pt plus3.5pt minus0pt
\def\@listi{\parsep 4.5pt plus 2pt minus 1pt
      \itemsep \parsep
      \topsep 9pt plus 3pt minus 3pt}}

\relax

\def\section{\@startsection{section}{1}{\z@}{3.5ex plus 1ex minus  .2ex}{2.3ex plus .2ex}{\large\bf}}

\def\thesection{\arabic{section}}
\def\thesubsection{\arabic{section}.\arabic{subsection}}

\def\appendix{\setcounter{section}{0}
 \def\thesection{Appendix \Alph{section}}
 \def\thesubsection{\Alph{section}.\arabic{subsection}}
 \def\theequation{\Alph{section}.\arabic{equation}}}
\renewcommand{\theequation}{\arabic{section}.\arabic{equation}}

}{
\renewcommand{\theequation}{\arabic{equation}}

} 

\begin{document}
\def\thefootnote{\fnsymbol{footnote}}
\def\thetitle{Evidence for the Unbinding of the $\phi^4$ Kink's Shape Mode}
\def\autone{Jarah Evslin}
\def\affa{Institute of Modern Physics, NanChangLu 509, Lanzhou 730000, China}
\def\affb{University of the Chinese Academy of Sciences, YuQuanLu 19A, Beijing 100049, China}

\title{Titolo}

\ifthenelse{\equal{\pr}{1}}{
\title{\thetitle}
\author{\autone}
\affiliation {\affa}
\affiliation {\affb}

}{

\begin{center}
{\large {\bf \thetitle}}

\bigskip

\bigskip


{\large \noindent  \autone{${}^{1,2}$}}


\vskip.7cm

1) \affa\\
2) \affb\\

\end{center}

}

\begin{abstract}
\noindent
The $\phi^4$ double-well theory admits a kink solution, whose rich phenomenology is strongly affected by the existence of a single bound excitation called the shape mode.  We find that the leading quantum correction to the energy needed to excite the shape mode  is $-0.115567\lambda/m$ in terms of the coupling $\lambda/4$ and the meson mass $m$ evaluated at the minimum of the potential.   On the other hand, the correction to the continuum threshold is $-0.433\lambda/m$.   A naive extrapolation to finite coupling then suggests that the shape mode melts into the continuum at the modest coupling of $\lambda/4\sim 0.106 m^2$, where the $\Z_2$ symmetry is still broken.

\noindent

\end{abstract}

%
\setcounter{footnote}{0}
\renewcommand{\thefootnote}{\arabic{footnote}}

\ifthenelse{\equal{\pr}{1}}{
\maketitle
}{}

\section{Introduction}
The $\phi^4$ double well field theory admits a kink solution.  This kink appears in a number of physical systems, from polyacetylene \cite{ssh} to crystals \cite{crystal} to graphene \cite{graphene}.  Besides the translation mode, it enjoys a single bound excitation, called the shape mode.  The shape mode is believed to be responsible for most of the kink's distinctive phenomenology \cite{campos2020} (but see Ref.~\cite{weigel2018}), such as the resonance phenomenon \cite{campres,quintero2000} (but see Ref.~\cite{takyi}),  wobbling kink multiple scattering \cite{alonso2020,campos2021} and spectral walls \cite{adamwall,wall2}.  While the shape mode deserves and has received significant attention in the literature, quantum corrections to the shape mode have not yet been considered.

In this note we report the one-loop quantum correction to the energies needed to excite both the kink's shape mode and also the first continuum excitation.  We find that the energy needed to excite a shape mode decreases linearly in the coupling, but that the energy needed to excite a continuum mode decreases more rapidly.  If this behavior is extrapolated to intermediate couplings, then the shape mode energy will exceed that of a continuum meson and so the shape mode is expected to become unbound, dissolving into the continuum at $\lambda/m^2\sim 0.422$.  This occurs at smaller coupling than other expected features of the model, such as the restoration of the $\phi\rightarrow -\phi$ symmetry at $\lambda/m^2\sim 1.2(1)$ found in Ref.~\cite{serone}, or the point at which the meson mass exceeds twice \cite{mussardo06,bajnok} or $\pi$ times \cite{slava15} the semiclassical \cite{dhn2,mephi4} kink mass and one expects mesons to drop out of the spectrum.

\section{Review of the $\phi^4$ Kink}

The $\phi^4$ double-well quantum field theory is defined by the Hamiltonian
\bea
H&=&\int dx \left[\frac{1}{2}:\pi(x)\pi(x):_a+\frac{1}{2}:\partial_x\phi(x)\partial_x\phi(x):_a+:\frac{\phi^2}{4}\left(\sqrt{\lambda}\phi(x)-m\sqrt{2}\right)^2:_a
\right]
\eea
where $::_a$ is normal ordering with respect to the operators that create plane wave excitations of the $\phi$ field about a classical vacuum.  It admits a classical kink solution
\beq
\phi(x,t)=f(x)=\frac{m}{\sqrt{2\lambda}}\left(1+\tanh\left(\frac{m x}{2}\right)\right) \label{f} 
\eeq
whose mass is classically
\beq
Q_0=\frac{m^3}{3\lambda}.
\eeq

Small, orthonormal perturbations about the kink solution with frequency $\omega_k$ will all be called normal modes.  They include a zero-mode proportional to $f\p$ as well as continuum modes $g_k(x)$ and a shape mode $g_\kt(x)$
\bea
g_k(x)&=&\frac{e^{-ikx}}{\ok{} \sqrt{m^2+4k^2}}\left[2k^2-m^2\right. \label{nmode}\\
&&\left.
+(3/2)m^2\sech^2\left(\frac{m x}{2}\right)-3im k\tanh\left(\frac{m x}{2}\right)\right]\nonumber\\
g_S(x)&=&-i\frac{\sqrt{3m}}{2}\tanh\left(\frac{m x}{2}\right)\sech\left(\frac{m x}{2}\right).\nonumber
\eea
At tree level the energy required to excite a normal mode is its frequency \cite{dhn2}
\beq
\omega_k=\sqrt{m^2+k^2}\hsp
\omega_\kt=\frac{\sqrt{3}}{2}m.
\eeq

\section{Energy Required to Excite a Shape Mode}

\begin{figure}
\setlength{\unitlength}{.55cm}
\begin{picture}(6,4)(-8,-1)

\put(21.5,-.5){\vl{-1}0 {1.25}}
\put(19,-.5){\vl{-1}0 {1.25}}
\put(20.25,-.2){\makebox(0,0){{$\kt$}}}
\put(17.75,-.2){\makebox(0,0){{$\kt$}}}
\put(19,.5){\circle{2}}

\put(11.5,2){\circle{2}}
\put(12.5,-1){\vector(-1,2){.5}}
\put(12.5,-1){\line(-1,2){1}}
\put(12.5,0){\makebox(0,0){{$k$}}}
\put(14,-1){\vl{-1}0 {.75}}
\put(12.5,-1){\vl{-1}0 {1.25}}
\put(13.1,-.7){\makebox(0,0){{$\kt$}}}
\put(11.1,-.7){\makebox(0,0){{$\kt$}}}

\put(7.5,2){\circle{2}}
\put(7.5,1){\vector(-1,-2){.5}}
\put(7.5,1){\line(-1,-2){1}}
\put(7.5,0){\makebox(0,0){{$k$}}}
\put(9,-1){\vl{-1}0 {1.25}}
\put(6.5,-1){\vl{-1}0 {0.75}}
\put(7.7,-.7){\makebox(0,0){{$\kt$}}}
\put(5.7,-.7){\makebox(0,0){{$\kt$}}}

\qbezier(-1,0)(0,1)(1,1)
\qbezier(-1,0)(0,0)(1,1)
\put(-.25,.625){\vector(-2,-1)0}
\put(.25,.375){\vector(-2,-1)0}
\put(-.6,.9){\makebox(0,0){{$k_1$}}}
\put(.5,.1){\makebox(0,0){{$k_2$}}}
\put(2,-1){\vector(-3,1){1.5}}
\put(2,-1){\line(-3,1)3}
\put(.5,-.9){\makebox(0,0){{$\kt$}}}
\put(1,1){\vector(-3,1){1.5}}
\put(1,1){\line(-3,1)3}
\put(-.3,1.8){\makebox(0,0){{$\kt$}}}

\put(-2,0){\vl{-1}0 {1}}
\put(-3,.3){\makebox(0,0){{$\kt$}}}
\put(-6,0){\vl{-1}0 {1}}
\put(-7,.3){\makebox(0,0){{$\kt$}}}
\qbezier(-4,0)(-5,1)(-6,0)
\qbezier(-4,0)(-5,-1)(-6,0)
\put(-5.1,.5){\vector(-1,0)0}
\put(-5.1,-.5){\vector(-1,0)0}
\put(-5,.9){\makebox(0,0){{$k_1$}}}
\put(-5,-.9){\makebox(0,0){{$k_2$}}}

\end{picture}
\caption{The first two diagrams give $\mui1$.  The next two are equal and sum to $\mui2$.  The last diagram is $\mui3$.  $\mui4$ is found by replacing one $k$ in the first two diagrams with a zero mode.  A loop factor $\I(x)$ arises for each loop at a single vertex.}
\label{mufig}
\end{figure}
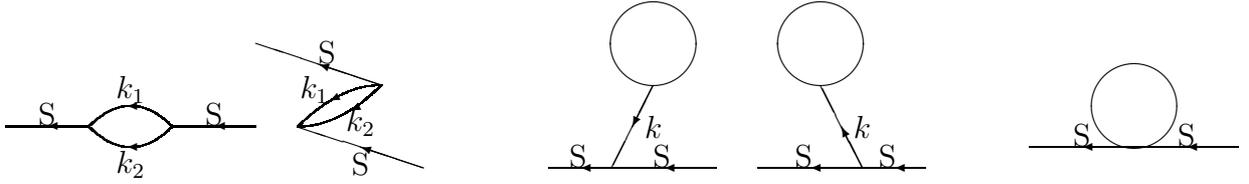

The one-loop energy required to excite the shape mode is \cite{menormal}
\bea
\mu(\kt)&=&\sum_{i=1}^4\mui{i}\hsp
\mui{1}=
\dpink{2}\frac{\left(\ok1+\ok2\right)\left|V_{k_1k_2\kt}\right|^2}{8\omega_\kt\ok1\ok2\left(\omega_\kt^2-\left(\ok1+\ok2\right)^2\right)}\nonumber\\
\mui{2}&=&-\dpin{k}\frac{V_{-\kt k \kt}V_{\I -k}}{4\omega_\kt\ok{}^2}\hsp
\mui{3}=\frac{V_{\I \kt -\kt}}{4\omega_\kt}\nonumber\\
\mui{4}&=&\frac{1}{4Q_0}\pin{k}\left(\frac{\omega_\kt}{\ok{}}+\frac{\ok{}}{\omega_{\kt}}
\right)\Delta_{-\kt-k}\Delta_{\kt k}.\label{muk}
\eea
The corresponding diagrams, defined in Ref.~\cite{menormal}, are drawn in Fig.~\ref{mufig}.  Here we have defined $\int^+ dk/(2\pi)$ to be an integral over all real values of $k$ plus the same expression with $k$ replaced by the shape mode.  The matrix $\Delta$ is defined by
\beq
\Delta_{Sk}=\int dx g_S(x) g\p_k(x)
\eeq
where it is understood that $g_{-S}=g^*_S$.

Let $V^{(n)}(x)$ be the $n$th functional derivative of $H$ with respect to $\phi(x)$, evaluated at $\phi(x)=f(x)$
\beq
V^{(3)}(x)=3\sqrt{2\lambda}m \tanh\left(\frac{m x}{2}\right)\hsp
V^{(4)}(x)=6\lambda.
\eeq
Then we have also defined the symbol
\beq
V_{k_1\cdots k_n}\,=\,\int\, dx V^{(n)}(x)\prod_{i=1}^n  g_{k_i}(x)
\hsp
V_{\I k_1\cdots k_n}\,=\,\int\, dx V^{(2+n)}(x)\I(x)\prod_{i=1}^n  g_{k_i}(x)
\label{vf}
\eeq
where the loop function $\I(x)$ is
\bea
\I(x)&=&\frac{\left|g_{\kt}(x)\right|^2}{2\omega_\kt}+\pin{k}\frac{\left(\left|g_{k}(x)\right|^2-1\right)}{2\omega_k} \label{di}\\
&=&\frac{1}{4\sqrt{3}}\sech^2\left(\frac{m x}{2}\right)\tanh^2\left(\frac{m x}{2}\right)-\frac{3}{8\pi}\sech^4\left(\frac{m x}{2}\right). \nonumber
\eea

Most of the relevant $x$ integrals were already evaluated analytically in Ref.~\cite{mephi4}
\bea
\Delta_{Sk}&=&i\pi \frac{\sqrt{3}}{16}\frac{(3m^2+4k^2)\sqrt{m^2+4k^2}}{m^{3/2}\ok{}}\sk\label{delta}\\
V_{SSS}&=&i\pi\frac{9\sqrt{3\lambda}}{32\sqrt{2}}m^{3/2}\hsp
V_{kSS}=i\pi\frac{3\sqrt{\lambda}}{\sqrt{2}}\frac{k^2\ok{}(m^2-2k^2)}{m^3\sqrt{m^2+4k^2}}\ck\nonumber\\
V_{k_1k_2S}&=&-i\pi \frac{3\sqrt{3\lambda}}{2\sqrt{2}}\frac{\left(\frac{17}{16}m^4-(k_1^2-k_2^2)^2\right)(m^2+4k_1^2+4k_2^2)+8m^2k_1^2k_2^2}{m^{3/2}\ok1\ok2\sqrt{m^2+4k_1^2}\sqrt{m^2+4k_2^2}}\sech\left(\frac{\pi(k_1+k_2)}{m}\right)\nonumber
\eea
including some terms with a loop factor $\I(x)$
\bea
V_{\I k}&=&i\frac{\sqrt{\lambda}}{\sqrt{6}}\frac{k^2\omega_k}{m^4\sqrt{m^2+4k^2}}\left[\pi(-m^2+2k^2)+3\sqrt{3}\omega_k^2\right]\ck\\
V_{\I S}&=&i\frac{3\sqrt{m\lambda}}{64\sqrt{2}}(3\sqrt{3}-2\pi).\nonumber
\eea
We will also need
\beq
V_{\I\kt-\kt}=\left(\frac{3\sqrt{3}\pi-18}{35\pi}
\right)\lambda.
\eeq

Substituting these expressions into (\ref{muk}) one can immediately evaluate the third and fourth terms
\bea
\mui{3}&=&\left(\frac{3\pi-6\sqrt{3}}{70\pi}
\right)\frac{\lambda}{m}\sim -0.00439962\frac{\lambda}{m}\\
\mui{4}&=&\frac{3\sqrt{3}\pi^2\lambda}{2^{11}m^7}\pin{k} \frac{(7m^2+4k^2)(3m^2+4k^2)^2(m^2+4k^2)}{\left(m^2+k^2\right)^{3/2}}\sks\nonumber\\
&\sim&0.27419991\frac{\lambda}{m}.\nonumber
\nonumber
\eea
The second term can be decomposed into the contributions $\mu^{(20)}$ and  $\mu^{(21)}$ in which the dummy index is the shape mode or a continuum mode respectively
\bea
\mu^{(20)}(\kt)&=&-\frac{V_{-\kt \kt \kt}V_{\I -\kt}}{4\omega^3_\kt}=\frac{9\pi}{2^{11}}(3\sqrt{3}-2\pi)\frac{\lambda}{m}\sim -0.01500739\frac{\lambda}{m}\\
\mu^{(21)}(\kt)&=&-\pin{k}\frac{V_{-\kt k \kt}V_{\I -k}}{4\omega_\kt\ok{}^2}\nonumber\\
&=&\frac{\pi\lambda}{4m^8}\pin{k}\frac{k^4(m^2-2k^2)}{m^2+4k^2}
\left[(3\sqrt{3}-\pi)m^2+(3\sqrt{3}+2\pi)k^2\right]
\cks\nonumber\\
&\sim& -0.00339318\frac{\lambda}{m}.
\nonumber
\eea

Finally we may decompose the first contribution into terms $\mu^{1I}$ with $I$ dummy indices running over continuous modes $k$
\bea
\mui{10}&=&
\frac{\left(2\omega_\kt\right)\left|V_{\kt\kt\kt}\right|^2}{8\omega^3_\kt\left(\omega_\kt^2-\left(2\omega_\kt\right)^2\right)}=-
\frac{\left|V_{\kt\kt\kt}\right|^2}{12\omega^4_\kt}=-
\frac{9\pi^2}{2^{9}}\frac{\lambda}{m}\sim
-0.17348914\frac{\lambda}{m}\\
\mui{11}&=&
2\pin{k}\frac{\left(\ok{}+\omega_\kt\right)\left|V_{kS\kt}\right|^2}{8\omega^2_\kt\ok{}\left(\omega_\kt^2-\left(\omega_\kt+\ok{}\right)^2\right)}\nonumber\\
&=&\frac{3\pi^2\lambda}{m^8}\pin{k}\frac{\left(2\ok{}+\sqrt{3}m\right)\ok{}k^4}{\left(3m^2-\left(\sqrt{3}m+2\ok{}\right)^2\right)}
\frac{(m^2-2k^2)^2}{m^2+4k^2}\cks
\nonumber\\
&\sim& -0.01112149\frac{\lambda}{m}
\nonumber\\
\mui{12}&=&
\pink{2}\frac{\left(\ok1+\ok2\right)\left|V_{k_1k_2\kt}\right|^2}{8\omega_\kt\ok1\ok2\left(\omega_\kt^2-\left(\ok1+\ok2\right)^2\right)}\nonumber\\
&=&\frac{9\sqrt{3}\pi^2\lambda}{8m^4}
\pink{2}\frac{\left(\ok1+\ok2\right)}{\ok1^3\ok2^3}\sech^2\left(\frac{\pi(k_1+k_2)}{m}\right)\nonumber\\
&&\times\frac{\left[\left(\frac{17}{16}m^4-(k_1^2-k_2^2)^2\right)(m^2+4k_1^2+4k_2^2)+8m^2k_1^2k_2^2\right]^2}{(m^2+4k_1^2)(m^2+4k_2^2)\left(3m^2-4\left(\ok1+\ok2\right)^2\right)}\sim -0.1823560(2)\frac{\lambda}{m}
.
\nonumber
\eea

Adding these seven contributions, we find that the one-loop correction to the energy cost of exciting the shape mode is
\beq
\mu(\kt)\sim -0.1155669(2)\frac{\lambda}{m}.
\eeq

\section{Continuum Threshold}

We have seen that the cost of exciting the shape mode decreases with increasing coupling.  However, this does not guarantee that the shape mode will remain in the spectrum, as the continuum threshold also depends on the coupling.  In Ref.~\cite{menormal} it was shown that at one-loop, the energy required to create a continuum excitation is the sum of the one-loop correction to the meson mass in the vacuum sector plus the nonrelativistic recoil kinetic energy of the kink.  The continuum threshold is the mass of the meson with no momentum, and so there is no recoil.  This means that the continuum threshold energy is simply equal to the one-loop mass correction in the $\phi^3$ theory with mass $m$ and potential $m\sqrt{\lambda/2}\phi^3$, corresponding to a classical vacuum, which can be calculated in old-fashioned perturbation theory or found in a textbook, or can heuristically be read from (\ref{muk}) by replacing $f(x)$ with the classical vacuum and the normal modes with plane waves and dividing by a normalization factor of $2\pi\delta(0)$.  Any method yields
\beq
\mu(0)=\frac{9m\lambda}{2}\pin{p}\frac{1}{\omega_p(m^2-4\omega_p^2)}=-\frac{\sqrt{3}}{4}\frac{\lambda}{m}\sim -0.433013\frac{\lambda}{m}.
\eeq

Including the leading order and one-loop contributions, the energy required to excite a continuum mode is then
\beq
m-\frac{\sqrt{3}}{4}\frac{\lambda}{m}
\eeq
and that required to excite a shape mode is
\beq
\frac{\sqrt{3}}{2}m-0.1155669(2)\frac{\lambda}{m}.
\eeq
These are equal when $\lambda/m^2\sim 0.422$, at which point one expects the shape mode to delocalize into oblivion, drastically affecting the phenomenology.  As the coupling constant is $\lambda/4$, this corresponds to a rather small value of the coupling and so it seems reasonable to speculate that the higher order corrections only displace but do not avoid this conclusion.

We have presented evidence that, at reasonably low but finite coupling, the $\phi^4$ kink loses its only bound excitation that is not a zero mode.  It is not clear from our calculation whether such behavior is generic for kinks, or indeed for solitons.  Recently, in Ref.~\cite{takyiphi8} (see also \cite{takyi20,lozano}), a number of new examples of such bound excitations have been found, and an efficient algorithm was described for generating them.  Our method can only be applied to stable kinks, as we require Hamiltonian eigenstates, but using the stability criterion in Refs.~\cite{tstabile,wstabile} one may easily filter examples.  Our formula (\ref{muk}) can then be applied to study the evolution of the bound modes.

\section* {Acknowledgement}

\noindent
JE is supported by the CAS Key Research Program of Frontier Sciences grant QYZDY-SSW-SLH006 and the NSFC MianShang grants 11875296 and 11675223.   JE also thanks the Recruitment Program of High-end Foreign Experts for support.

\end{document}